\documentstyle[11pt,newpasp,twoside,epsf]{article}
\def\edcomment#1{\iffalse\marginpar{\raggedright\sl#1\/}\else\relax\fi}
\reversemarginpar

\begin{document}
\title{Recombination rate coefficients for astrophysical applications
from  storage-ring experiments}
 \author{S. Schippers, S. B\"{o}hm, and A. M\"{u}ller}
\affil{Institut f\"{u}r Kernphysik, Strahlenzentrum der Justus-Liebig-Universit\"{a}t, 35392 Giessen,
Germany}
\author{G. Gwinner, M. Schnell, D. Schwalm, and A. Wolf}
\affil{Max-Planck-Institut f\"{u}r Kernphysik, 69029 Heidelberg, Germany}
\author{D. W. Savin}
\affil{Columbia Astrophysics Laboratory, Columbia University, NY 10027, USA}

\begin{abstract}
The basic approach for measuring electron-ion recombination rate coefficients in merged-beams
electron-ion collision experiments at heavy-ion storage rings is outlined. As an example
experimental results for the low temperature recombination of C\,{\sc iv} ions are compared with
the recommended theoretical rate coefficient by Mazzotta et al. The latter deviates by factors of
up to 5 from the experimental one.
\end{abstract}

\section{Introduction}

One important process that governs the charge state balance in a plasma is dielectronic
recombination (DR). Accordingly, DR rate coefficients form a basic ingredient in plasma modeling
codes that are employed for the analysis of spectra, e.\,g., obtained from astrophysical
observations. In order to be able to infer a reliable description of the plasma properties from
such calculations, accurate rate coefficients for the basic atomic collision processes in a plasma
are required. To date, most DR rate coefficients used for plasma modeling stem from theoretical
calculations. The calculation of DR rate coefficients is a challenging task since an infinite
number of states is involved in this process. Approximations and computational simplifications are
needed in order to make DR calculations tractable. It turns out that different calculations yield
rate coefficients differing by up to orders of magnitude. In this situation benchmarking
experiments are vitally needed in order to guide the development of the theoretical methods and to
provide accurate DR rate coefficients for plasma modelers.

In the past decade electron-ion collision experiments employing merged beams at heavy-ion storage
rings have proven to be the primary approach for obtaining accurate experimental DR rate
coefficients (M\"{u}ller \& Wolf 1997; M\"{u}ller\& Schippers 2001). Several fundamental issues have been
addressed in this research such as the high resolution spectroscopy of dielectronic resonances, the
interference between radiative and dielectronic recombination and the influence of external
electromagnetic fields on DR (for recent concise reviews see Schippers 1999; Schippers et al.\
2000). Moreover, the electron-impact ionization of highly charged ions can also be studied at
heavy-ion storage rings. From the measured cross sections plasma rate coefficients are readily
derived as has been done, for example, for Fe\,{\sc xvi} (M\"{u}ller 1999).

In order to meet astrophysical needs, a dedicated experimental program on DR of L-shell iron ions
is currently being carried out at the heavy-ion storage ring {\sc TSR} of the Max-Planck-Institut
f\"{u}r Kernphysik in Heidelberg, Germany. Already the first results for Fe\,{\sc xviii}, Fe\,\,{\sc
xix} (Savin et al.\ 1997; Savin et al.\ 1999) revealed distinct differences between experimental
and theoretical DR rate coefficients, especially at low temperatures where these ions may form in
photoionized plasmas. Results for more highly charged ions ranging from Fe\,{\sc xx} (Savin et al.\
2001) up to Fe\,{\sc xxiv} have also been obtained already and will be published in the near
future. In the following, the state of affairs is exemplified by comparing the measured C\,{\sc iv}
DR rate coefficient (Schippers et al.\ 2001) with the recent recommendation by Mazzotta et al.\
(1998).

\section{Outline of the Experimental Procedure}

\begin{figure}
\plotfiddle{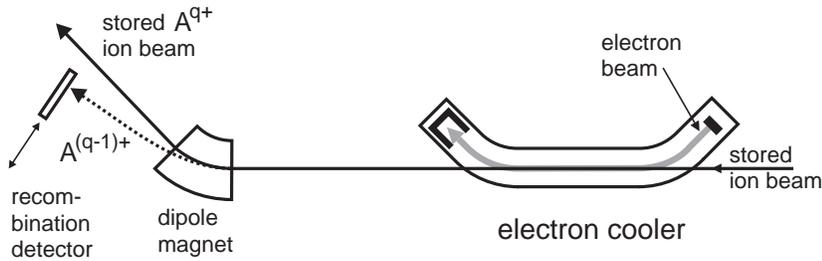}{3.3cm}{0}{60}{60}{-190pt}{-205pt} \caption{Sketch of the experimental
setup. The beam of ions with parent charge state $q$ enters the electron cooler from the right. In
the bending magnet behind the electron cooler recombined A$^{(q-1)+}$ ions are bent less than the
circulating parent A$^{q+}$ ions. After charge state separation the recombined ions are counted by
a single-particle detector with nearly 100\% efficiency.}
\end{figure}

\begin{figure}
\plotfiddle{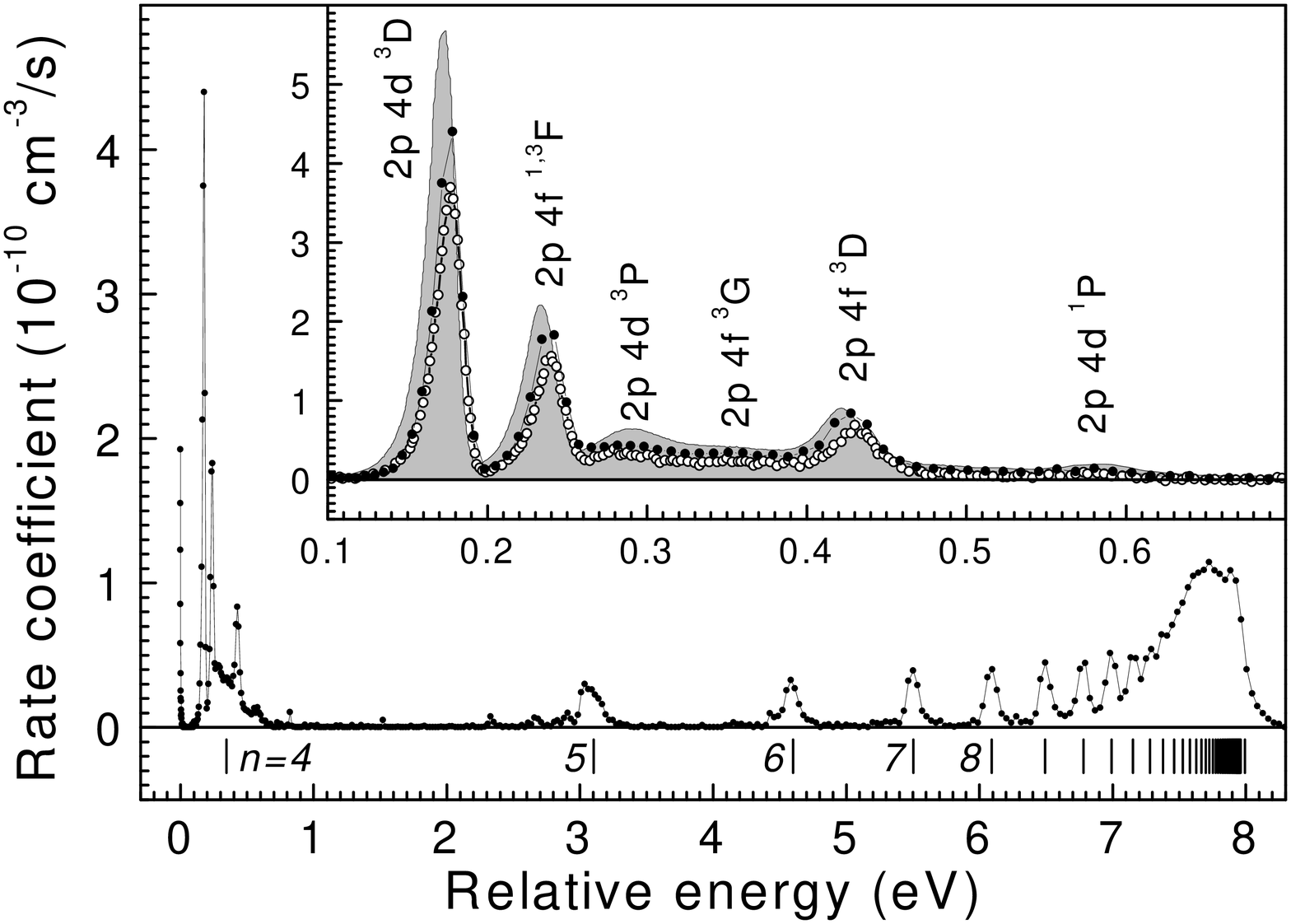}{6cm}{0}{30}{30}{-130pt}{-0pt} \caption{Experimental C\,{\sc iv}
merged-beams rate coefficient (full circles, Schippers et al.\ 2001). The vertical dashes denote
the $2pnl$ resonance positions according to a Rydberg formula .  The inset shows the  C\,{\sc iv}
DR spectrum in the region of the $2p\,4l$ resonances, the experimental {\sc cryring} data (open
circles), and the theoretical results (shaded curve) of Mannervik et al.\ (1998). Their peak
designations are given for the most prominent DR resonances.}
\end{figure}

Electron-ion recombination experiments at a heavy-ion storage ring is conceptually simple. An ion
beam with a unique charge-to-mass ratio from an accelerator is injected into the storage ring. By
using multiple injection and electron-cooling techniques ion currents up to a few hundred $\mu$A or
even a few mA can be accumulated in the heavy-ion storage ring {\sc TSR} of the Max-Planck-Institut
f\"{u}r Kernphysik in Heidelberg. Electron cooling is the transfer of the ion beam's internal kinetic
energy to a heat bath consisting of cold electrons. It is most effective when the electrons move
with nearly the same average velocity as the ions. This condition is achieved in the collinear part
of the interaction region inside the electron cooler, inserted into one of the straight sections of
the storage ring (see Fig.\ 1). In the electron cooler an electron beam is merged with the ion
beam, guided along the ion beam and finally demerged by toroidal and axial magnetic fields. Since
the stored ions move with typically 10\% of the speed of light electron energies of about 3~keV are
required in order to reach the cooling condition.

As a result of electron cooling the ion beam shrinks to a diameter of typically 2~mm and its
internal velocity spread is reduced to below 0.01\%. Moreover, after the cooling period of
typically a few seconds, any metastable states that initially might have been present in the ion
beam have vanished. The resulting extremely well defined ion beam is subsequently used for
recombination experiments where the electron cooler is used as an electron target. Non-zero
relative energies $E_{\rm rel}$ between electrons and ions are introduced by detuning the electron
energy from its value required for cooling. In order to prevent internal ion beam heating from
becoming effective, the detuning periods only last a few milliseconds, after which cooling is
restored. Absolute ($\pm 15\%$ uncertainty) merged-beams recombination rate coefficients as a
function of $E_{\rm rel}$ are obtained from the measured recombination count rate on the
single-particle recombination detector (Fig.\ 1) by appropriate normalization on the simultaneously
measured ion and electron currents.

\section{Results and Discussion}

\begin{figure}
\plotfiddle{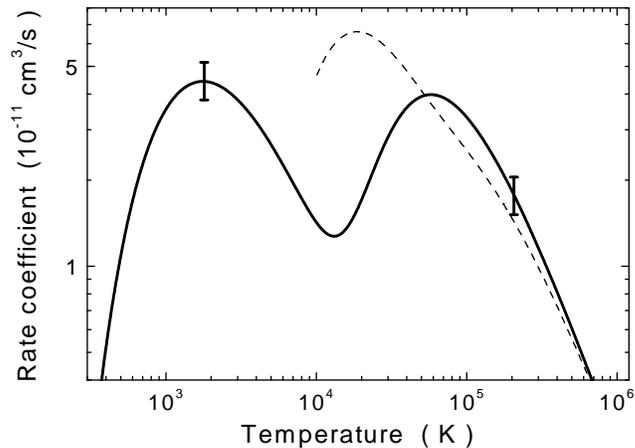}{6cm}{0}{30}{30}{-130pt}{-0pt} \caption{ C\,{\sc iv} DR rate
coefficients in a plasma. Thick full line: experimentally derived result of Schippers et al.\
(2001, systematic uncertainty $\pm 15\%$), thin dashed line: recommendation by Mazzotta et al.\
(1998).}
\end{figure}

As an example Fig.~2 shows the C\,{\sc iv} merged-beams recombination rate coefficient as a
function of $E_{\rm rel}$ (Schippers et al.\ 2001). Due to the low experimental energy spread,
which is mainly determined by the electron beam temperature (here 0.15 meV and 10 meV in the
longitudinal and transverse degrees of freedom, respectively), individual $2p\,nl$ DR resonances
associated with $2s\to2p$ core excitations are resolved for $n\leq 9$. Toward the series limit
overlapping high-$n$ resonances produce the prominent structure around $E_{\rm rel} \approx
7.5$~eV. As discussed explicitly by Schippers et al.\ (2001) one experimental limitation is the cut
off of high Rydberg resonances in the motional electric field generated in the charge analyzing
dipole magnet (cf., Fig.~1). In the C\,{\sc iv} experiment the associated experimental cut off has
been around the quantum number $n_{\rm cut}=19$ and the unmeasured DR resonance strength
accumulated in resonances with $n>n_{\rm cut}$ has been substantial. This issue is much less severe
for more highly charged ions such as  Fe\,{\sc xviii} (Savin et al.\ 1997; 1999) where due to the
higher nuclear charge the DR resonance strengths decrease more rapidly with increasing $n$ and,
moreover, the experimental cut off occurs at a much higher $n_{\rm cut}$ typically larger than 100.

The inset of Fig.~2 enlarges the region of the $2p\,4l$ resonances where other experimental results
measured at the Stockholm heavy ion storage ring {\sc cryring} (Mannervik et al.\ 1998) are also
available. The agreement between the two experimental data sets is within the experimental
uncertainties. Mannervik et al.\ (1998) also performed theoretical calculations of the
recombination rate coefficient within the framework of relativistic many-body perturbation theory.
From these calculations it became clear that relativistic effects decisively determine the DR
resonance strength even for such a light ion as C\,{\sc iv}. Conclusively, it can be stated that
accurate theoretical recombination rate coefficients can probably only be obtained from
relativistic fine structure calculations and that simpler LS-coupling calculations are bound to be
in error.

For the derivation of the C\,{\sc iv} plasma recombination rate coefficient the measured
merged-beams recombination rate coefficient was first extrapolated to high $n$ by a theoretical
calculation in order to account for the unmeasured high-$n$ DR resonance strength. In a second step
the combined measured and extrapolated rate coefficient was convoluted by a Maxwellian electron
energy distribution characterized by the plasma electron temperature $T_{\rm e}$. The resulting
experimentally derived C\,{\sc iv} plasma DR rate coefficient (Schippers et al.\ 2001) is plotted
in Fig.~3 (thick full line) as a function of $T_{\rm e}$. Also shown in Fig.~3 is the recommended
C\,{\sc iv} DR rate coefficient by Mazzotta et al. (1998, thin dashed line). At temperatures below
50\,000 K it deviates by factors of up to 5 from our experimental result. When comparing with all
available theoretical C\,{\sc iv} DR rate coefficients  we find that none of them agrees with the
experimentally derived rate coefficient over the entire relevant temperature range (Schippers et
al.\ 2001).

In conclusion, heavy-ion storage rings are the tools of choice for the experimental determination
of absolute recombination rate coefficients. By comparison with the experimental results we find
that theoretical DR rate coefficients currently used in plasma modeling can be in serious error. In
view of these deficiencies we are presently carrying out an experimental program for the
determination of low temperature recombination rate coefficients of the iron L-shell ions Fe\,{\sc
xvii} to Fe\,{\sc xiv} (Savin et al.\ 1997; 1999; 2001). In the future we plan to extend these
measurements to higher excitation energies that are relevant for high temperature DR occuring in
the solar as well as in stellar coronae.

Note added in proof: A recent theoretical treatment of C\,{\sc iv} recombination has been published by A. K.
Pradhan et al. 2001, Phys. Rev. Lett. 87, 183201. 
\end{document}